\documentclass{ws-ijmpa}
\usepackage[super,compress]{cite}
\usepackage{graphicx}
\usepackage{float}
\usepackage[nodisplayskipstretch]{setspace}
\begin{document}
\markboth{SDG, A. Ranjan, H. Nandan}{Regge Trajectories of Tetraquarks and Pentaquarks}

%
\catchline{}{}{}{}{}
%

\title{Regge Trajectories of Tetraquarks and Pentaquarks with Massive Quarks in the Flux Tube Model}

\author{Sindhu D G$^{1*}$, Akhilesh Ranjan$^{1\dagger}$ and Hemwati Nandan$^{2,3,4\ddagger}$}
\address{$^{1}$Department of Physics, Manipal Institute of Technology, Manipal Academy of Higher Education, Manipal, Karnataka 576104 India.\\$^{2}$Department of Physics, Hemwati Nandan Bahuguna Garhwal Central University, Srinagar Garhwal, Uttarkhand 246174, India.\\ $^{3}$Department of Physics, Gurukula Kangri (Deemed to be University), Haridwar, Uttarakhand 249404, India.\\ $^{4}$Center for Space Research, North-West University, Mafikeng 2745, South Africa.
\\$^{*}$sindhudgdarbe@gmail.com\\$^{\dagger}$ak.ranjan@manipal.edu\\$^{\ddagger}$hnandan@associates.iucaa.in}
\maketitle
\begin{history}
\received{Day Month Year}
\revised{Day Month Year}
\end{history}
\begin{abstract}
In recent years, many tetraquarks and pentaquarks have been discovered by various 
experimental groups and $X(3872)$, $Z_{c}(3900)$, $X(4430)$, $P_{c}^{+}(4312)$, 
$P_{c}^{+}(4457)$ are some of the interesting observed tetraquark and pentaquark states. 
The Regge trajectories of some such states are studied in view of the flux tube model of hadrons with finite quark masses. The effect of  flux tube (or string) length variation on the Regge trajectories of these sates is analysed in detail. 
It is observed that for a fixed angular momentum, the string length has a  constant value. Some other states are also proposed and the results obtained are then compared with the studies by others. Our findings correspond rather well with those of other researchers and with those of the experiment.

\end{abstract}

\ccode{PACS numbers: 11.55.Jy. 12.40.Yx }
\section{Introduction}	
Each quark flavour in nature comprises a colour degree of freedom, according to quantum chromodynamics (QCD) and is represented by  the colour gauge group  $SU(3)_C$. The hadrons i.e.  baryons (with three quarks) and the mesons (with quark and an antiquark) belong to the singlet representation of the $SU(3)_C$ group. This demonstrates that the hadrons that have been seen are all colour singlets. As a result, these quarks in nature are restricted to the hadron's interior, making it impossible to detect a single quark in nature. Since every hadron is a colour singlet, we can anticipate the existence of more colour singlet multiquark systems. Gellmann was the first to put forward this idea.\cite{gellmann} Both tetraquark and pentaquark systems will be taken into account in this study. Tetra and pentaquark systems  are hadrons made up of four or five quarks, respectively. These systems were not discovered until the last century, and many of their characteristics are still unknown because of their high mass and short life time. 
The existence of tetraquark and pentaquark system has recently been supported by a number of experimental studies as discussed below.\\
Recently many experimental findings have confirmed the existence of tetraquarks and pentaquarks. In 2003, the Belle experiment in Japan discovered a tetraquark state $X(3872)$,\cite{choi} later confirmed by BABAR Collaboration, \cite{aubert} the Collider Detector at Fermilab experimental collaboration (CDF II),\cite{acosta,aaltonen} LHCb (Large Hadron Collider beauty),\cite{aaij1} D0 experiment \cite{abazov} and CMS (Compact Muon Solenoid).\cite{chatrchyan} Later many new tetraquark states like  $Y(4140)$ at the Belle collaboration in 2009, \cite{shen} $Z_{c}(3900)$ at BESIII in 2013 were also discovered \cite{ablikim}. In July 2015, in the decays of J/$\psi$  baryons, LHCb established the existence of two pentaquark states $P_{c}^{+}$(4380), $P_{c}^{+}$(4450) with the quark content ($uudc\bar{c}$). \cite{aaij2} Later other pentaquark 
candidates $P_{c}^{+}(4312)$, $P_{c}^{+}(4457)$, and pentaquark with strange quark, $P_{cs}(4459)$ were discovered. \cite{aaij3,aaij4} Theoretically, different studies have predicted the properties of fully charm tetraquark states, \cite{jing,zhi,chen,bedolla,jin,tommy} and in 2020, LHCb 
collaboration reported its experimental observation. \cite{aaij5} Inspired by this discovery, fully heavy pentaquark states were also studied. \cite{zang,hong} The very recent discovery is the tetraquark state with two heavy quark flavours and two light antiquark flavours such as $cc\bar{u}\bar{d}$ and $cc\bar{s}\bar{s}$ is worth noticing. \cite{aaij6,gao} 
In order to comprehend the nature of the strong force and colour confinement, it is crucial to analyse the features of these tetraquark and pentaquark states. To investigate the characteristics of these extremely confined quarks, many models are applied. A hadron's mass and angular momentum are intimately correlated by studies on Regge trajectories.\cite{guidry,cheng} The Regge trajectories of hadrons have been studied using a variety of theoretical models. Olsson string model, Soloviev string quark model, non-relativistic quark models like Inopin model, Martin model, relativistic and semi-relativistic models like Semay model, relativistic flux tube model, etc. are some example of theoretical models.\cite{olsson,soloviev,inopin1,inopin2,martin,semay1,semay2} To investigate the mass spectrum and Regge trajectories of various hadrons, flux tube models are utilised.\cite{semay2,bing}
Olsson and colleagues discovered the precise and almost classical solutions to the generalised Klein-Gordon (KG) problem by taking into account the concepts of string theory.\cite{olsson}
Using a composite fermion model of quasi-particles, diquarks are explained. Additionally, they have examined the Regge trajectories of the charm and bottom pentaquarks. Charm pentaquark's Regge trajectory was discovered to be linear, whereas the bottom pentaquark exhibits a deviation. These pentaquarks with angular momentum $J$ have ground state masses that have been determined and are in good agreement with the findings of other literature.\cite{ghosh} For mesons and baryons, Nandan {\it et al}\cite{nandan1,nandan2} updated the Regge trajectory equations.
While baryonic Regge trajectories displayed non-linear behaviour, the mesonic Regge trajectories were discovered to be linear.\\  
This paper is organized as follows. In section 2,  the expressions for classical mass and angular momentum are derived from the flux tube model for different tetraquark configurations. The obtained results are then discussed  in length in section 3. The conclusions drawn presented accordingly in section 4. The effect of string length variation on some of the tetraquark and pentaquark states are investigated in this paper and the determined masses of these states are compared with the experimental and other theoretical data.  The observed non linear behaviour of Regge trajectories for tetraquark states is worth noticing.

\section{Classical Mass and Angular Momentum of Tetraquark States}
\label{II}
\begin{figure}[H]
\begin{center}
\includegraphics[scale=0.4]{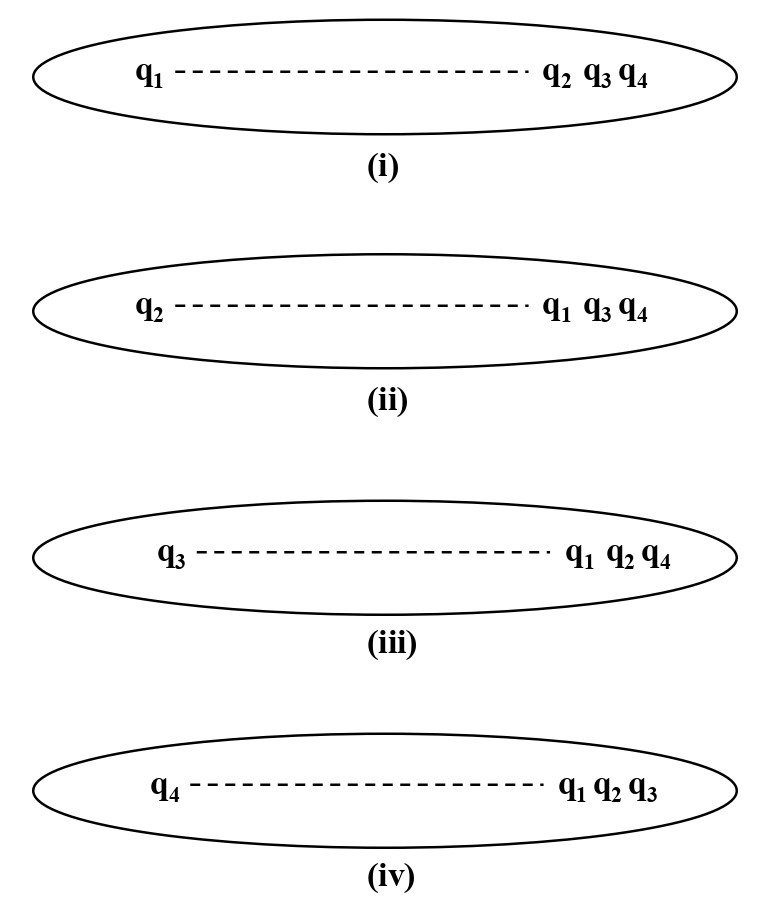}
\caption{Different configurations of tetraquarks with one quark at one end of the string.}
\label{fig:1}
\end{center}
\end{figure}

\begin{figure}[H]
\begin{center}
\includegraphics[scale=.4]{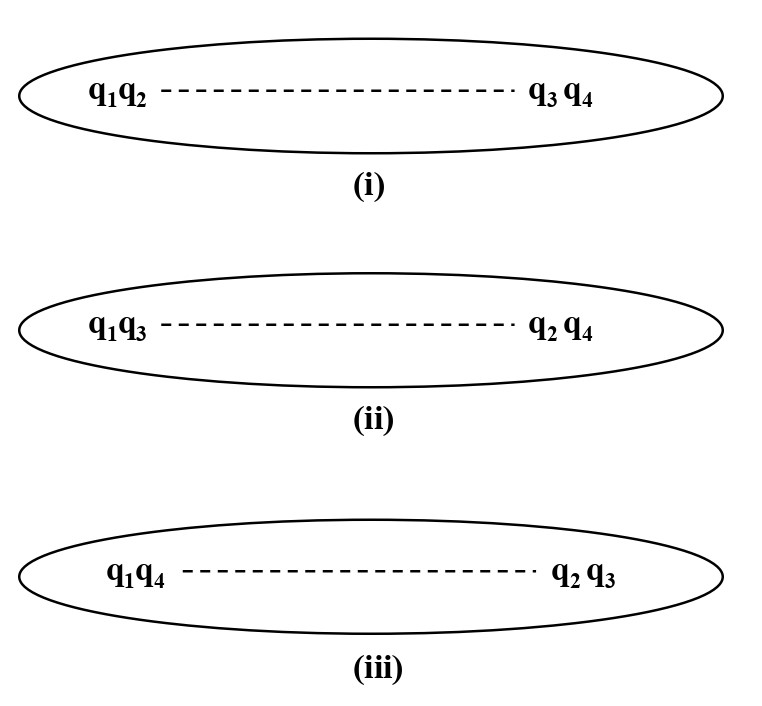}
\end{center}
\caption{Different configurations of tetraquarks with two quarks at the end of the string.}
\label{fig:2}
\end{figure}

The endpoints of the string in the flux tube model of hadrons, where the massless quarks are located at the string's end, are considered to revolve at the speed of light. Due to the string tension, the hadron's mass subsequently becomes its potential energy. The relationship between the classical mass and angular momentum of a hadron is given as follows if the string is rotating about its middle point:
\begin{equation} 
J=\alpha_{0}+\alpha M^{2}
\label{appeqn}
\end{equation}
where $\alpha_{0}$ and $\alpha$ are constants 
with $\alpha$=1/(2$\pi K$). Here K is the string tension. 
The behaviour of  Regge trajectories of hadrons is marked by above relation with the linear confining potential of the form, $V(r) = Kr$, where $r$ is the inter quark distance of the massless quarks lying at the ends of the string. The mass and angular momentum expressions for hadrons and pentaquarks are modified  and the Regge trajectories of the same are analysed accordingly. 
The present work is extended for the case of tetraquarks. In the flux tube model, for the case of tetraqurk, there will be seven different configurations as shown in Fig. \ref{fig:1} and \ref{fig:2}. The Fig. \ref{fig:1} shows the 
set of configurations of tetraquarks with one quark at one end of the string. The modified expression  for mass and angular momentum corresponding to the first configuration of Fig. \ref{fig:1} is as shown:\cite{nandan1,nandan2}
\begin{equation}
\begin{aligned}
M_{mod}&=
\frac{K(M-m_{q_{1}})l}{fM}\left(\int_{0}^{f}\frac{dv}{\sqrt{1-v^{2}}}+\int_{0}^{\frac{m_{q_{1}}}{M-m_{q_{1}}}f}\frac{dv}{\sqrt{1-v^{2}}}\right)\\
&+\gamma_{\alpha}m_{q_{1}}+\gamma_{\beta}(M-m_{q_{1}}).
\end{aligned} 
\label{appeqn}
\end{equation}
\setstretch{1}
\begin{equation}
\begin{aligned}
J_{mod}&=\frac{kl^{2}}{f
^{2}}\times \left(\frac{M-m_{q_{1}}}{M}\right)^{2} \left\lbrace \int_{0}^{f}\frac{v^{2}dv}{\sqrt{1-v^{2}}}+\int_{0}^{\frac{m_{q_{1}}}{M-m_{q_{1}}}f}\frac{dv}{\sqrt{1-v^{2}}}\right\rbrace\\
& +\frac{m_{q_{1}}fl}{M-m_{q_{1}}}\left\lbrace \gamma_{\alpha}(M-m_{q_{1}})+\gamma_{\beta}m_{q_{1}}\right\rbrace.
\end{aligned} 
\label{appeqn}
\end{equation}
Here, $\gamma_{\alpha}=\frac{1}{\sqrt{1-f^{2}}}$, $\gamma_{\beta}=\frac{1}{\sqrt{1-\left(\frac{m_{q_{1}}f}{M-m_{q_{1}}}\right)^{2}}}$ and $M=m_{q_{1}}+m_{q_{2}}+m_{q_{3}}+m_{q_{4}}$ is the sum of all individual quark masses and $f$ is the fractional 
rotational speed (actual speed is $fc$ with $c$=1 in natural system of units).\\
Equations (2) and (3) are displayed below after integration:
\begin{equation}
\begin{aligned}
M_{1i}=
\frac{K(M-m_{q_{1}})l}{fM}\left\lbrace\sin^{-1}f+\sin^{-1}\left(\frac{m_{q_{1}}f}{M-m_{q_{1}}}\right)\right\rbrace+\gamma_{\alpha}m_{q_{1}}+\gamma_{\beta}(M-m_{q_{1}}).
\end{aligned} 
\label{appeqn}
\end{equation}

\begin{equation}
\begin{aligned}
J_{1i}&=\frac{kl^{2}}{f
^{2}}\times \left(\frac{M-m_{q_{1}}}{M}\right)^{2} \left\lbrace \frac{1}{2}\sin^{-1}f-\frac{f}{2}\sqrt{1-f^{2}}+
\frac{1}{2}\sin^{-1}\left(\frac{m_{q_{1}}f}{M-m_{q_{1}}}\right)\right.\\
&\left.-\frac{fm_{q_{1}}}{2(M-m_{q_{1}})}
\times\sqrt{1-\left(\frac{m_{q_{1}}f}{M-m_{q_{1}}}\right)^{2}}\right\rbrace +\frac{m_{q_{1}}fl}{M-m_{q_{1}}}\left\lbrace \gamma_{\alpha}(M-m_{q_{1}})+\gamma_{\beta}m_{q_{1}}\right\rbrace.
\end{aligned} 
\label{appeqn}
\end{equation}

The Fig. \ref{fig:2} shows the set of tetraquark configurations with two 
quarks at the two ends of the string. 
The modified expression for mass and angular momentum corresponding to the first configuration of Fig. \ref{fig:2} is as shown:
\begin{equation}
\begin{aligned}
M_{mod}&=
\frac{K(m_{q_{3}}+m_{q_{4}})l}{fM}\left(\int_{0}^{f}\frac{dv}{\sqrt{1-v^{2}}}+\int_{0}^{\frac{m_{q_{1}}+m_{q_{2}}}{m_{q_{3}}+m_{q_{4}}}f}\frac{dv}{\sqrt{1-v^{2}}}\right)\\
&+\gamma_{\alpha}\left(m_{q_{1}}+m_{q_{2}}\right)+\gamma_{\beta}(m_{q_{3}}+m_{q_{4}}).
\end{aligned} 
\label{appeqn}
\end{equation}
\begin{equation}
\begin{aligned}
J_{mod}&=\frac{kl^{2}}{f
^{2}}\times \left(\frac{m_{q_{3}}+m_{q_{4}}}{M}\right)^{2} \left\lbrace \int_{0}^{f}\frac{v^{2}dv}{\sqrt{1-v^{2}}}+\int_{0}^{\frac{m_{q_{1}}+m_{q_{2}}}{m_{q_{3}}+m_{q_{4}}}f}\frac{dv}{\sqrt{1-v^{2}}}\right\rbrace\\
&+\frac{m_{q_{1}}+m_{q_{1}}}{m_{q_{3}}+m_{q_{4}}}fl\left\lbrace\gamma_{\alpha}\left(m_{q_{3}}+m_{q_{4}}\right)+\gamma_{\beta}\left(m_{q_{1}}+m_{q_{2}}\right)\right\rbrace.
\end{aligned} 
\label{appeqn}
\end{equation}
Equations (6) and (7) are presented here after integration:
\begin{equation}
\begin{aligned}
M_{2i}&=
\frac{Kl}{fM}(m_{q_{1}}+m_{q_{2}})+\left\lbrace\sin^{-1}f+\sin^{-1}\left(\frac{m_{q_{1}}+m_{q_{2}}}{m_{q_{3}}+m_{q_{4}}}f\right)\right\rbrace\\
&+\gamma_{\alpha}(m_{q_{1}}+m_{q_{2}})+\gamma_{\beta}(m_{q_{3}}+m_{q_{4}}).
\end{aligned}
\label{appeqn}
\end{equation}

\begin{equation}
\begin{aligned}
J_{2i}&=
\frac{Kl^{2}}{f^{2}}\frac{(m_{q_{3}}+m_{q_{4}})^{2}}{M^{2}}
\left\lbrace \frac{1}{2}\sin^{-1}f-\frac{f}{2}\sqrt{1-f^{2}}+\frac{1}{2}sin^{-1}\left(\frac{m_{q_{1}}+m_{q_{2}}}{m_{q_{3}}+m_{q_{4}}}f\right)\right.\\
&\left.-\left(\frac{m_{q_{1}}+m_{q_{2}}}{m_{q_{3}}+m_{q_{4}}}\right)\frac{f}{2}\times\left\lbrace 1-\left(\frac{m_{q_{1}}+m_{q_{2}}}{m_{q_{3}}+m_{q_{4}}}f\right)^{2}\right\rbrace^{\frac{1}{2}}\right\rbrace+\frac{m_{q_{1}}+m_{q_{2}}}{m_{q_{3}}+m_{q_{4}}}fl\\
&\times\left\lbrace\gamma_{\alpha}(m_{q_{3}}+m_{q_{4}})+\gamma_{\beta}(m_{q_{1}}+m_{q_{2}})\right\rbrace.
\end{aligned}
\label{appeqn}
\end{equation}
Here, $\gamma_{\beta}=\frac{1}{\sqrt{1-\left(\frac{m_{q_{1}}+m_{q_{2}}}{m_{q_{3}}+m_{q_{4}}}f\right)^{2}}}$. 
We assume that all the seven tetraquark configurations have equal probability
to occur, therefore, the actual mass and angular momentum must be averaged. As 
$\sin\theta \le 1$ $\implies f\le \frac{M-m_{q_{1}}}{m_{q_{1}}}$ (corresponding to equation (2)). 
From the special theory of relativity, $f\le1$. Hence these conditions 
must be satisfied.\\
Similarly, there can be fifteen different possible configurations for pentaquarks. There will be ten configurations with two and three quarks at the extreme ends of the string in the flux tube model, as well as five variants with one quark at one end and four quarks at the other (see Fig. 1 and Fig. 2 in Nandan {\it et al} \cite{nandan2} for complete detail).
\section{Discussion and Results}
\label{III}
\begin{figure}[H]
\label{fig:3}
\begin{center}
\includegraphics[scale=.35]{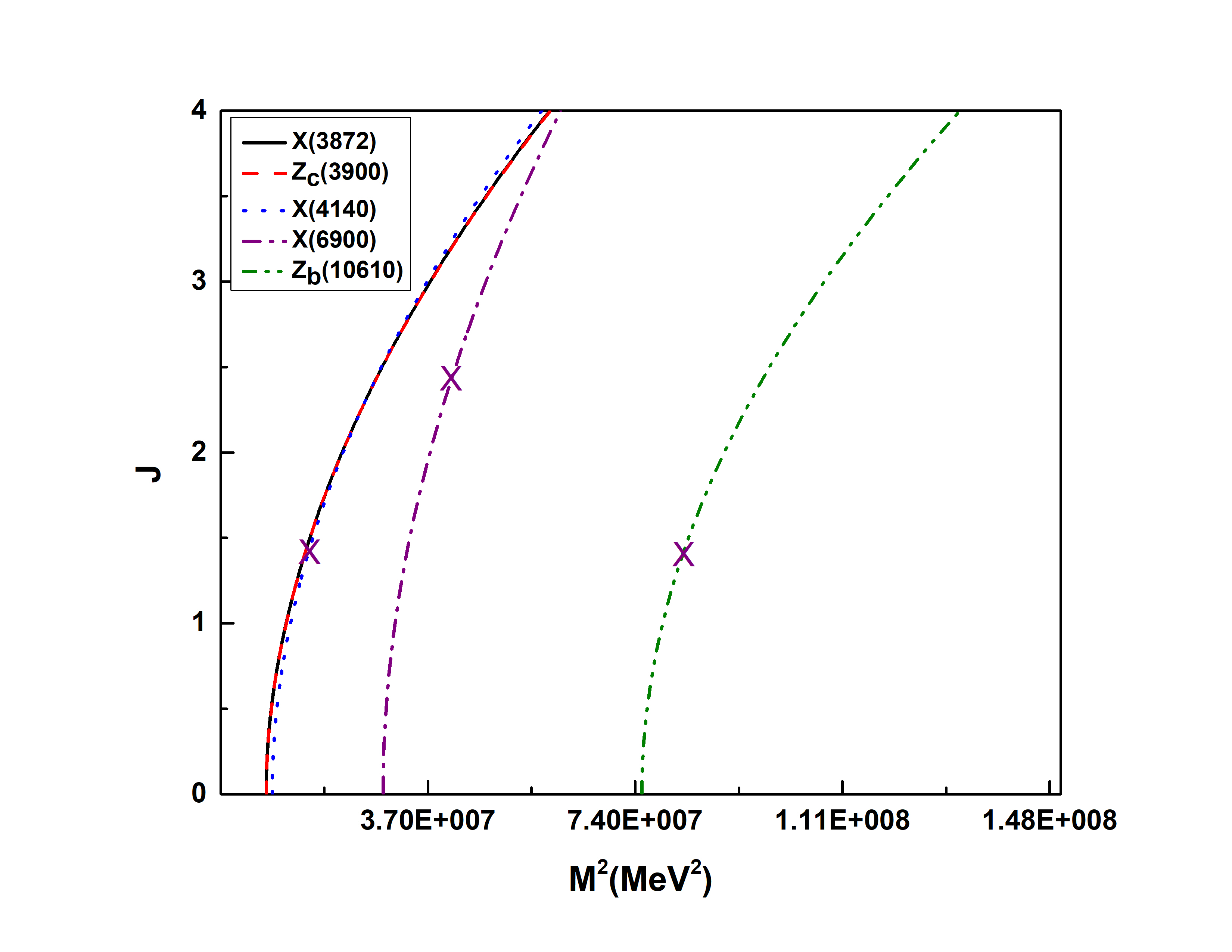}
\caption{Regge trajectories for different tetraquarks.}
\end{center}
\end{figure}

The masses of up, down, strange, charm, and bottom quarks 
considered for calculation are, $m_{u}=2.16MeV$, $m_{d}=4.67MeV$, 
$m_{s}=93MeV$, $m_{c}=1270MeV$ and $m_{b}=4180MeV$ 
respectively,\cite{pdg1} and $K=0.2GeV^{2}$.\cite{pdg2} 
The tetraquark mass variation rises linearly with string length. As shown by the calculated expression, it is obvious. The mass will increase together with the inter-quark distance as the system's energy does as well. We can anticipate additional quark pairs if this energy exceeds to some level. Table \ref{ta2} is more transparent for pentaquark states. The mass adjustment changes depending on the quark makeup, being more for heavy quarks and smaller for light quarks (see table \ref{ta3}). As a result, we can conclude that the QCD effects that lead to mass corrections get stronger with quark mass.
On the other hand, it is clearly evident from Table \ref{ta1} that light quark hadrons have a longer string. If we switch to heavy quark hadrons, it steadily drops. The mass correction, however, rises. Consequently, it may be concluded that the QCD effects that result in mass corrections grow with quark mass. The percentage mass correction, however, actually goes down from light to heavy quark flavours if we look at Table \ref{ta3}.
Higher states exhibit increased confinement and kinetic effects as a result of the decreasing string length and rising string end rotational speed. On the other hand, Table \ref{ta3} shows that if we move to heavier flavour states, both the confinement effect and the kinetic impact would co-decrease. Pentaquark masses are computed in Table \ref{ta2} and the results are compared with the experimental findings, which show a good level of agreement. However, the $\Delta M$ values for the pentaquark states are the same as those in Table \ref{ta3}. It demonstrates that $l$ and $f$ have no bearing on the mass correction. Table \ref{ta4} lists some projected novel tetraquark states, and the findings are contrasted with those of prior theoretical studies. The Regge trajectories of a few tetraquark states are depicted in Fig. \ref{fig:3}. It is evident that the angular momentum sharply rises with mass.
Tetraquarks' Regge trajectories are incredibly nonlinear. For various states, the string length $l$ becomes different. Fig. \ref{fig:3} displays the Regge trajectories of several pentaquark states. We observe that it is not linear. But it seems to be linear for higher order levels. Since the variations in mass with string length and mass with speed for tetraquarks are identical to those for pentaquarks \cite{nandan2}, these are not explicitly shown here.
Table \ref{ta1} lists the masses of several tetraquark states and contrasts them with the findings of the experiment. For various states, we have taken into account various $l$ values. The current findings and the outcomes of the experiment are well-aligned. According to the Table \ref{ta1}, string length $l$ reduces as the mass of the tetraquark rises. It demonstrates that the confinement effect increases with increasing mass states.  
\begin{figure}[H]
\label{fig:4}
\begin{center}
\includegraphics[scale=.35]{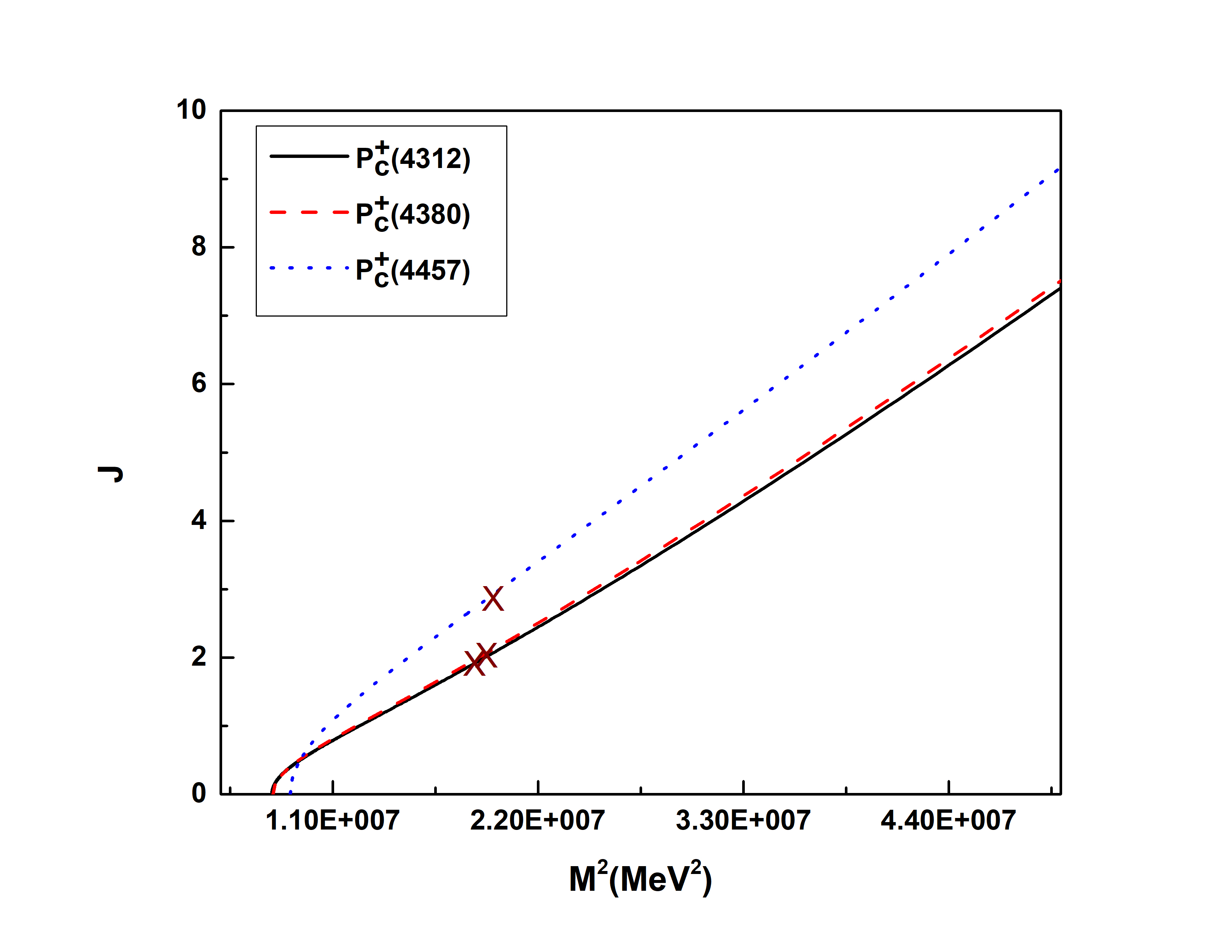}
\caption{Regge trajectories for different pentaquarks.}
\end{center}
\end{figure}

\begin{table}[H]
\tbl{Comparison of the results with the experimental results (tetraquarks).}
{\begin{tabular}{|c|c|c|c|c|c|c|}
\hline \noalign{\smallskip} 
S. No.&State&Quark&$J$&$M_{cal}$&$f$&$l$\\
&&structure&&($MeV$)&&($fm$)\\
\noalign{\smallskip}
\hline
1&$X(3872)$&$c\bar{c}u\bar{u}$&1&3874.38&0.761&0.302\\
\hline
2&$Z_{c}(3900)$&$c\bar{c}u\bar{d}$&1&3889.17&0.764&0.3\\
\hline
3&$X(4140)$&$c\bar{c}s\bar{s}$&1&4150&0.786&0.265\\
\hline
4&$X(4430)$&$c\bar{c}d\bar{u}$&1&4434.96&0.884&0.195\\
\hline
5&$X$(6900)&$c\bar{c}c\bar{c}$&2&6877&0.746&0.24\\
\hline
6&$Z_{b}(10610)$&$b\bar{b}u\bar{d}$&1&10599.75&0.672&0.127\\
\hline
7&$Z_{b}(10650)$&$b\bar{b}u\bar{d}$&1&10657&0.678&0.125\\
\hline
\end{tabular}
\label{ta1}}
\end{table}

\begin{table}[H]
\tbl{Comparison of the results with the experimental results (pentaquarks).}
{\begin{tabular}{|c|c|c|c|c|c|c|}
\hline \noalign{\smallskip} 
S. No.&State&Quark&$J$&$M_{cal}$&$f$&$l$\\
&&Configuration&&($MeV$)&&($fm$)\\
\noalign{\smallskip}
\hline
1&$P_{c}^{+}(4312)$&$c\bar{c}uud$&$\frac{3}{2}$&4324.2&0.848&0.23\\
\hline
2&$P_{c}^{+}(4380)$&$c\bar{c}uud$&$\frac{3}{2}$&4372.34&0.854&0.22\\
\hline 
3&$P_{c}^{+}(4457)$&$c\bar{c}uud$&$\frac{5}{2}$&4444.94&0.839&0.4\\
\hline
\end{tabular}
\label{ta2} }
\end{table}

\begin{table}[H]
\tbl{Mass correction fraction for tetraquarks and pentaquarks.}
{\begin{tabular}{|c|c|c|c|c|}
\hline \noalign{\smallskip} 
S. No.&State&Quark&$\Delta M$&$\frac{\Delta M}{M}$\\
&&Configuration&($MeV$)&\\
\noalign{\smallskip}
\hline
1&$X(3872)$&$c\bar{c}u\bar{u}$&1327.68&0.343\\
\hline
2&$Z_{c}(3900)$&$c\bar{c}u\bar{d}$&1353.17&0.347\\
\hline
3&$X(4140)$&$c\bar{c}s\bar{s}$&1414&0.342\\
\hline
4&$X(4430)$&$c\bar{c}d\bar{u}$&1883.17&0.425\\
\hline
5&$X$(6900)&$c\bar{c}c\bar{c}$&1820&0.264\\
\hline
6&$Z_{b}(10610)$&$b\bar{b}u\bar{d}$&2243.17&0.211\\
\hline
7&$Z_{b}(10650)$&$b\bar{b}u\bar{d}$&2283.17&0.214\\
\hline
8&$P_{c}^{+}(4312)$&$c\bar{c}uud$&1763&0.409\\
\hline
9&$P_{c}^{+}(4380)$&$c\bar{c}uud$&1831&0.418\\ 
\hline
10&$P_{c}^{+}(4457)$&$c\bar{c}uud$&1908&0.428\\
\hline
\end{tabular}
\label{ta3} }
\end{table}

\begin{table}[H]
\tbl{Comparison of the results with other studies.}
{\begin{tabular}{|c|c|c|c|c|c|c|}
\hline \noalign{\smallskip} 
S No.&Quark&$J$&$M_{cal}$&$f$&$l$&Other\\
&Configuration&&($MeV$)&&($fm$)&Studies; \cite{wang,qi}\\
\noalign{\smallskip}
\hline
1&$ss\bar{s}\bar{s}$&1&2029.96&0.988&0.303&$2210\pm 90$\\
\hline
2&$us\bar{s}\bar{c}$&1&3170.18&0.951&0.5&3156\\
\hline
3&$ss\bar{s}\bar{c}$&1&3383.78&0.956&0.41&3379\\
\hline
4&$us\bar{s}\bar{b}$&1&6433.82&0.864&0.64&6464\\
\hline
\end{tabular}
\label{ta4} }
\end{table}

\section{Conclusions}
The major objective of the current work is to examine how string length affects the Regge trajectories of the tetraquark and pentaquark states. The string lengths are all within a 1fm range. By taking into account the current quark masses, the Regge trajectories of tetraquarks and pentaquarks are determined. And by employing these recent quark masses, the masses produced are well characterised. We can also see from the results that the string length gets shorter with heavier states. When compared to tetraquark states, the pentaquark states' Regge trajectories are practically linear. When the rotational speed is low, the masses of the tetraquark states are linear, but as the speed rises, they become significantly nonlinear. In order to generate the higher angular momentum states of tetraquarks and pentaquarks, more experimental data is required. 
\section*{Acknowledgments}
AK is thankful to Manipal Academy of Higher Education (MAHE) Manipal for the 
financial support under scheme of intramural project grant no. 
MAHE/CDS/PHD/IMF/2019. SDG is thankful to  `Dr. T. M. A. Pai Scholarship 
Program' for the financial support.










\end{document}